# 1. GNSS Jamming Detection with Automatic Gain Control (AGC) and Carrier-to-Noise Ratio Density (CNO) Observables from a COTS receiver

Syed Ali Kazim[1], Anas Darwich[2], Juliette Marais[3]

1 - IRT Railenium, 59540 Valenciennes, France
2 - SNCF, 2 PLACE AUX ETOILES 93210 SAINT-DENIS, France
3 - Univ Gustave Eiffel, COSYS-LEOST, F-59650 Villeneuve d'Ascq, France

## 1.1. Introduction

As rail transport moves toward higher degrees of automation under initiatives like the R2DATO project [1], accurate and reliable train localization has become essential. Global Satellite Navigation System (GNSS) is considered as a main technology in enabling operational advancements including Automatic Train Operation (ATO), moving block signaling, and virtual coupling, which are the core components of the Horizon Europe 2024 rail digitalization agenda. However, GNSS signal integrity is increasingly threatened by intentional and unintentional radio frequency interference (RFI). This include jamming and spoofing, which are particularly concerning as the broadcasted signal can deliberately disrupt or manipulate the GNSS signal.

- *Jamming* refers to an intentional form of interference that induces disturbances in the GNSS band, causing performance degradation or can even entirely block the receiver from acquiring the satellite signals.
- *Spoofing* involves broadcasting counterfeit satellite signals to deceive the GNSS receiver, leading to inaccurate estimation of position, navigation and timing information.

This concern about interference is not unique to rail applications. The aeronautical sector has long recognized the risks posed by GNSS interference, with extensive documentation on its impact on navigation, landing procedures, and surveillance systems. In recent years, awareness of these risks has expanded to other transport sectors. Within the automotive industry, particularly in Intelligent Transport Systems (ITS), several studies [2][3][4] have addressed the vulnerability of GNSS against interference. Similar concerns are now emerging in the rail domain [5][6][7], especially as GNSS is increasingly adopted in safety-critical applications.

In literature, several levels of actions have been explored, ranging from merely the detection of a malicious signal at the initial phase to the application of advanced signal processing methods aimed at suppressing the effects of interference [8]. In alignment with the goal of the R2DATO project, we evaluated the impact of various classes of interference signals such as amplitude modulation (AM), frequency modulation (FM), pulsed, frequency hopping and chirp signals on the GNSS observables including Automatic Gain Control (AGC) and Carrier to Noise Ratio (CNO) as measured by a Commercial Off-The-Shelf (COTS). However, in this work, the analysis is only limited to impact of chirp interference on GPS L1 receiver observables and detection performance.

## 1.2. Interference signal and detection observables

According to the survey [8], different types of jamming signals have been reported. Each type poses a unique challenge to reliable signal processing because of their distinct characteristics. However, as mentioned earlier, this study specifically focuses on frequency-modulated or chirp-like signals.

### 1.2.1. Linear chirp interference

A linear chirp signal is a type of frequency-modulated waveform in which the instantaneous frequency increases or decreases linearly over time. The chirp interference signal can be expressed as,

$$i(t) = \sqrt{P_I}\, e^{j(f_{inst}(t)t + \theta_I)} \quad [1.1]$$
$$f_{inst}(t) = 2\pi f_I + \pi b t\, (f_{max} - f_{min})/T_{sweep} \quad [1.2]$$

where, $f_I$ denotes the starting frequency, $b$ represents the sweeping direction, which can be either upward ($b = +1$) or downward ($b = -1$), $T_{sweep}$ is the sweep period, and $f_{max} - f_{min}$ is the sweep bandwidth with and $f_{min}$ and $f_{max}$ represent the minimum and maximum frequency range. Figure 1. Illustrate the time frequency representation of a linear chirp signal.

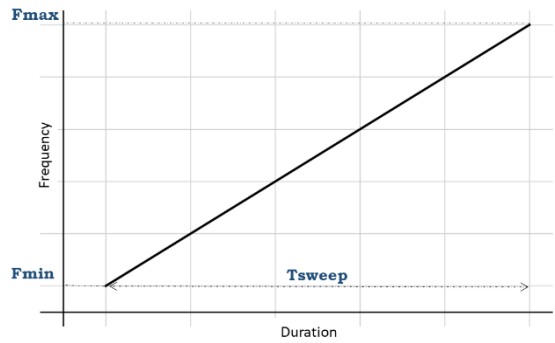

**Figure 1. Time-Frequency representation of linear chirp.**

### 1.2.2. Automatic Gain Control (AGC)

AGC is a critical component in the GNSS receiver front-end that maintains consistent signal levels by adjusting its gain to keep the signal level within the ADC's dynamic range. Under normal conditions, AGC operates steadily near the thermal noise level due to the weak nature of the signal. However, when interference increases the input signal power, AGC rapidly reduces gain to avoid ADC saturation and this sudden drop of gain could indicate the presence of interference. Interference detection using AGC is based on monitoring the deviation from the nominal level in interference-free conditions by comparing it with the threshold, which can be expressed as follows,

$$Th_{AGC} = \mu_{ref} - 3\sigma_{ref} - T_{drop} \quad [1.3]$$

where, $\mu_{ref}$ is the mean value and $\sigma_{ref}$ is the standard deviation of interference-free samples. The drop threshold $T_{drop}$(= 2dB) to compensate for the natural fluctuations in AGC values. The interference flag is triggered when the observed value drops below the expected nominal level.

### 1.2.3. Carrier to Noise Ratio (CNO)

CNO is an important parameter indicating the signal strength of the received GNSS signal. It quantifies the ratio of received carrier power to the noise density, typically expressed in dB-Hz. For each satellite channel, CNO can be estimated as [9],

$$CNO = 10\, log_{10}(\frac{1}{T}\frac{\hat{\mu}_{NA}-1}{M-\hat{\mu}_{NA}}) \quad [1.4]$$

where, $T$ is the code duration in seconds (i.e., 0.001 s for the GPS L1 C/A signal), $M$ is the total number of $T$ blocks used for coherent integration (usually $M = 20$ as the data bit duration for the GPS L1 C/A signal is 20 ms), and $k$ is the number of samples used to compute average normalized power $\hat{\mu}_{NA}$, which can be expressed as follows:

$$\hat{\mu}_{NA} = \frac{1}{K}\sum_{k=1}^{K}\frac{\left(\sum_{i=1}^{M} I_{p_i}\right)_k^2 + \left(\sum_{i=1}^{M} Q_{p_i}\right)_k^2}{\left(\sum_{i=1}^{M}(I_{p_i}^2 + Q_{p_i}^2)\right)_k} \quad [1.5]$$

where, $I_{p_i}$ and $Q_{p_i}$ are the prompt correlator output at the tracking stage from in-phase and quadrature component, respectively. The jamming signal induces distortions by increasing the noise level, causing drop in the CNO values across multiple satellites. The detection method implemented monitors CNO variations concurrently and triggers an interference event when the observed CNO from several satellites simultaneously falls below a predefined threshold (= 5dB) level.

1.3. Experimental setup

The schematic diagram with different tools and software used in this work is shown in Figure 2. It includes an IQ replay system, a customizable jamming signal generator and a Septentrio (AsteRx SBi3) receiver to acquire GNSS measurements specifically, CNO and AGC gain from the MeasEpoch and ReceiverStatus blocks respectively.

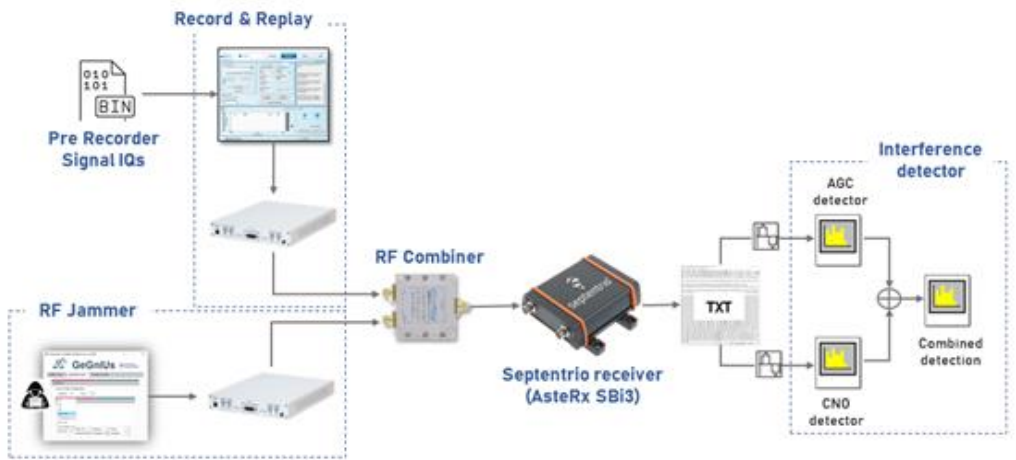

**Figure 2. Experimental setup used for GNSS interference detection.**

In this configuration, the pre-recorded IQ samples collected by SNCF along the railway line between Auterive and Pamiers in southwestern France are replayed on the GPS L1 frequency. At regular intervals, the chirp signal of 30-second duration is generated and combined with the GNSS signal through RF combiner, as depicted in Figure 3. On the receiver end, data is continuously logged for 30 minutes, during which at each interference interval the jamming signal strength is increased by 5dB. The logged data is then post-processed and employed in the development phase.

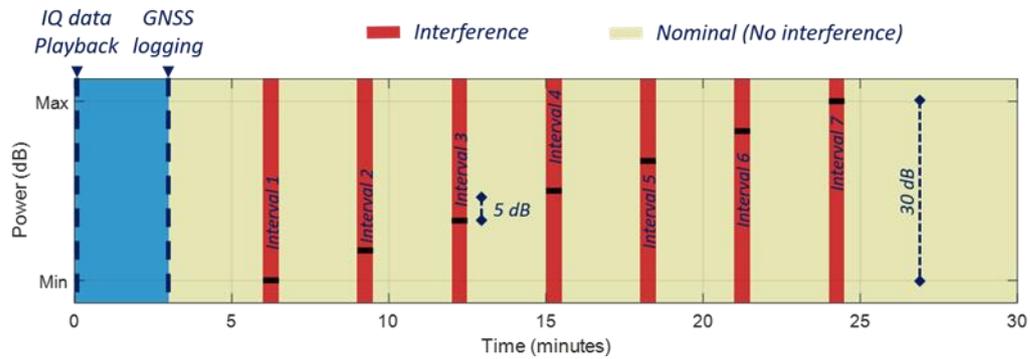

Figure 3. GNSS playback scenario including jamming in multiple intervals

1.4. Results and Analysis

As a representative example, Figure 4 shows the effects of the chirp signal on AGC and CNO observables. When interference is present, AGC lowers the receiver gain to keep the signal at a nominal level. The adaptive response becomes more evident within each successive interval as interference power is increased by 5dB. As the interference power increases, the AGC reduces its gain even more to compensate for additional power in the GNSS band. Similarly, the CNO indicating signal quality also dropped as interference induces distortion in the useful signal. During the final interference interval, it appears that the receiver loses tracking of the satellites when maximum power is applied, resulting in missing data during this period.

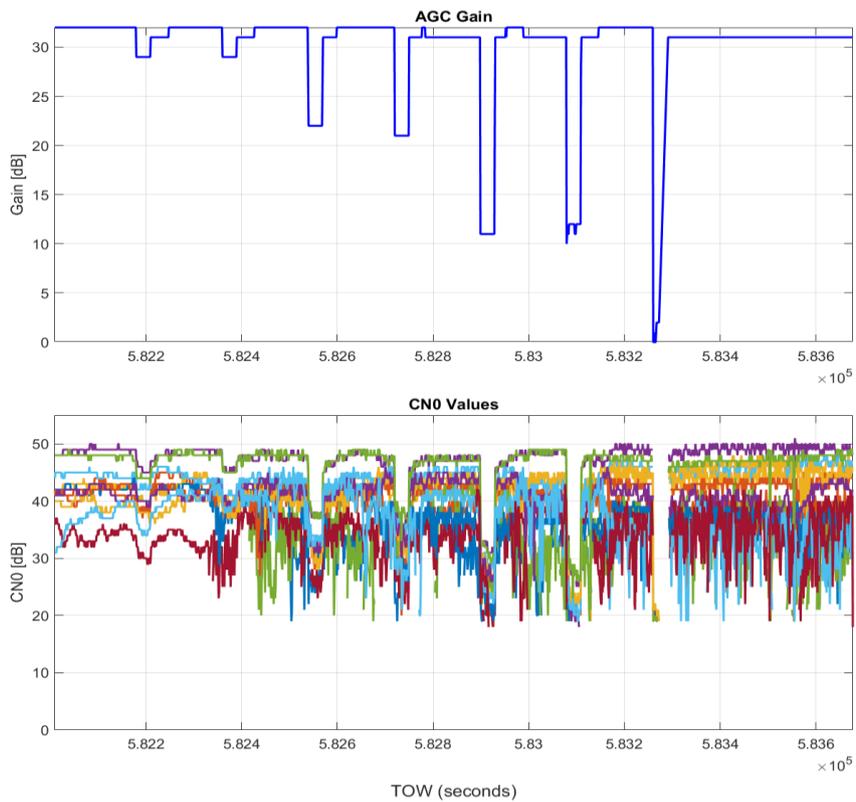

Figure 4. Linear Chirp with a) AGC response and b) CNO of GPS L1 frequency

In the following, the performance of the interference detection algorithm is evaluated using AGC and CNO observables. Figure 5 presents the results obtained from recorded data, with detection performed using a) CNO-based detector (upper panel) and b) AGC-based detector (lower panel). The AGC-based detector consistently raised detection flags across all interference intervals (1-7). In contrast, the CNO-based detector could not detect low-power interference, particularly in intervals 1 and 2. However, it successfully identifies events in intervals 3-7 as the jamming power is increased.

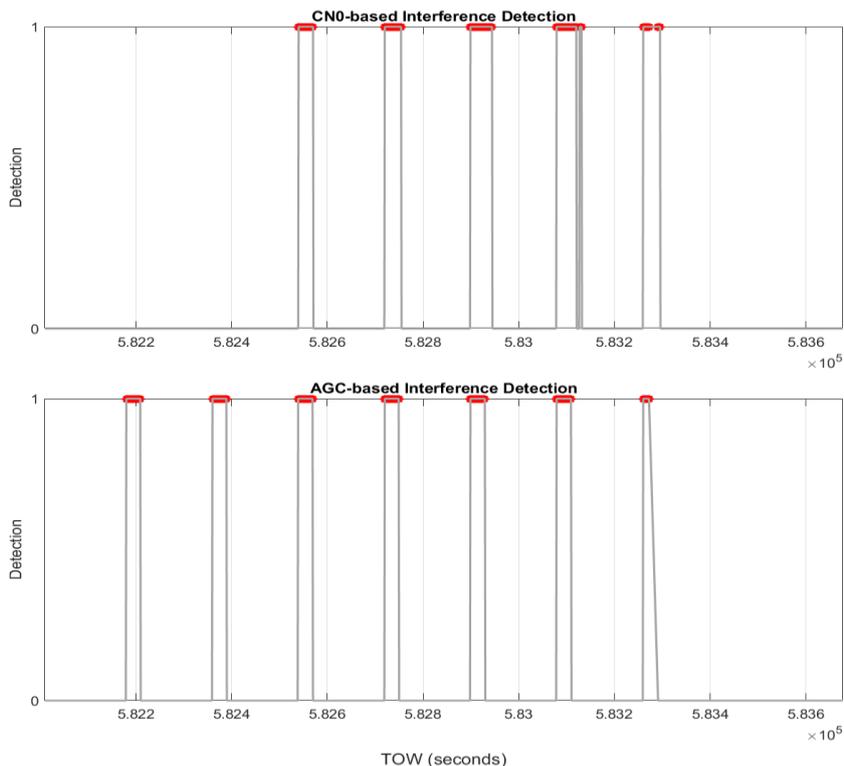

Figure 5. Linear Chirp detection using a) AGC-based detector, b) CNO-based detector.

Table 1 summarizes the detection results. Here, AGC response serves as the reference for the expected detections. The results show that the CNO-based detector achieved a detection rate of 76.5% with a 23% false alarm rate. Here, the false detection seems likely due to the settling time taken by the tracking loop filters during the recovery phase before the satellite signals are re-acquired and restored to their nominal state.

Table 1. Detection results summary.

|  | AGC-based detector | CNO-based detector |
|---|---|---|
| Interference detected intervals | 7/7 | 5/7 |
| Detection probability | 100% | 76.5% |
| Missed detection rate | 0% | 0.5% |
| False alarm rate | 0% | 23.0% |

1.5. Conclusions

This study demonstrates that AGC and CNO observables from COTS receivers offer a practical and effective approach for GNSS jamming detection. However, each metric has inherent limitations when used independently. AGC, which is primarily sensitive to signal power, can also be influenced by temperature variations, which were not considered in our analysis. On the other hand, environmental conditions, satellite geometry, and multipath effects can also affect CNO values. These factors can reduce detection reliability in certain conditions. However, considering both AGC and CNO observables could however be used to enhance the detection sensitivity and overall robustness.

Bibliography

[1] https://www.railjournal.com/in_depth/r2dato-accelerating-the-digitalisation-and-automation-of-europes-railway-network/
[2] R. Bauernfeind, T. Kraus, A.S. Ayaz, D. Dötterböck and B. Eissfeller. Analysis, detection and mitigation of incar GNSS jammer interference in intelligent transport systems. Deutsche Gesellschaft für Luft-und Raumfahrt-Lilienthal-Oberth eV, 2013.
[3] S. Dasgupta, K.H. Shakib, and M. Rahman. "Experimental Validation of Sensor Fusion-based GNSS Spoofing Attack Detection Framework for Autonomous Vehicles". In: arXiv preprint arXiv:2401.01304 (2024).
[4] S. Kocher, J. Hansen, and A. Rügamer. "GNSS Interference Localization for Vehicular Jammers using low-cost COTS Sensors". In: 2024 International Conference on Localization and GNSS (ICL-GNSS). IEEE. 2024, pp. 1–7.
[5] R. Ehrler, A. Wenz, S. Baumann, P. Mendes, N. Dütsch, A. Martin and C. Hinterstocker. "Jamming and spoofing impact on GNSS signals for railway applications". In: Proceedings of the 36th International Technical Meeting of the Satellite Division of The Institute of Navigation (ION GNSS+ 2023). 2023, pp. 4153–4167.
[6] S. Wang, J. Liu, B.G. Cai, J. Wang and D.B. Lu "BOSVDD-based GNSS Spoofing Detection for Rail Vehicle Positioning". In: IEEE Transactions on Instrumentation and Measurement (2024).
[7] A. Vennarini, A. Coluccia, D. Gerbeth, O.G. Crespillo and A. Neri. "Detection of gnss interference in safety critical railway applications using commercial receivers". In: Proceedings of the 33rd International Technical Meeting of the Satellite Division of The Institute of Navigation (ION GNSS+ 2020). 2020, pp. 1476–1489.
[8] R. Morales-Ferre, P. Richter, E. Falletti, A. De La Fuente and E.S. Lohan. "A survey on coping with intentional interference in satellite navigation for manned and unmanned aircraft." IEEE Communications Surveys & Tutorials 22.1 (2019): 249-291.
[9] J.J. Spilker Jr, P. Axelrad, B.W. Parkinson, and P. Enge. eds., 1996. Global positioning system: theory and applications, volume I. American Institute of Aeronautics and Astronautics.

List of Abbreviations

ITS – Intelligent Transport Systems
COTS – Commercial Off-The-Shelf
AGC – Automatic Gain Control
CNO – Carrier to Noise Ratio

Zoom Authors

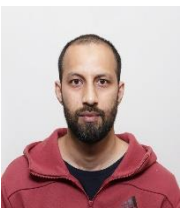

**Syed Ali Kazim** received the PhD degree in Signal and Image Processing from Université Gustave Eiffel, France, in 2024. He is currently working as a Research Engineer at IRT Railenium, where he contributes in the development of advanced solutions to enhance the resilience of GNSS-based localization systems. His research interests include integrity monitoring, fault detection algorithms and interference countermeasures with a focus on jamming signal.
syed.ali-kazim@railenium.eu

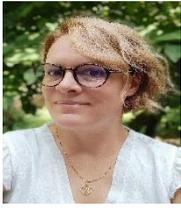 **Juliette Marais** received the engineering degree from ISEN and a Ph.D. degree in electronics from University of Lille, France, in 1998 and 2002 respectively. Since 2002, she has been a Researcher with INRETS, IFSTTAR and now University Gustave Eiffel. She is currently involved in GNSS performance analyses and enhancement in land transport environments and in particular in rail applications. Since 2000, she has been participating with the European Railway-Related Projects, such as the recent H2020 RAILGAP project or supporting EUSPA with expertise. Her research focuses the development of fail-safe positioning solutions for land transport applications and GNSS propagation characterization in railway environments (NLOS, Multipath, Interferences) including propagation phenomena, positioning and pseudorange error modelling, filtering technics, and simulation. With Railenium, she contributes to the R2DATO project.
Juliette.marais@univ-eiffel.fr

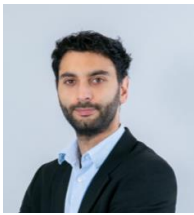 **Anas DARWICH** received an MSc in Navigation and Telecommunication from ENAC, France, in 2018. From 2018 to 2022, he worked at Safran Electronics & Defense as a systems engineer in aeronautical applications. Since 2023, he has been working at SNCF in the railway sector, where he participated in projects such as X2RAIL5 and CLUG2.0. Recently, he has been working on the R2DATO project, focusing on integrating navigation prototypes on trains, designing test platforms, and assessing navigation algorithm performance.

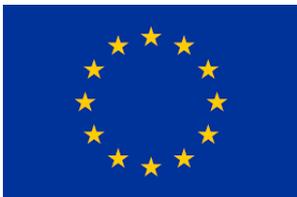
Funded by the European Union

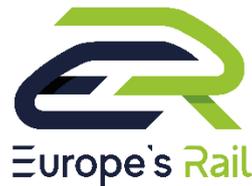
The project is supported by the Europe's Rail Joint Undertaking and its members. Funded by the European Union. Views and opinions expressed are however those of the author(s) only and do not necessarily reflect those of the European Union or the Europe's Rail Joint Undertaking. Neither the European Union nor the granting authority can be held responsible for them.